\documentstyle[12pt]{article}
\begin{document}
\title{Black hole entropy without brick walls }
\author{Li Xiang\footnote{e-mail: xiang.lee@163.com} \\
CCAST (World Lab.), P.O.Box, 8730, Beijing 100080,
\\People's Republic of China\\and\\ Institute of Theoretical Physics,
Chinese Academy of Sciences,
\\P.O.Box 2735, Beijing 100080, People's Republic of
China\footnote{this is the mailing address}}
\date{}
\maketitle
\begin{abstract}
The properties of the thermal radiation are discussed by using the
new equation of state density motivated by the generalized
uncertainty relation in the quantum gravity. There is no
burst at the last stage of the emission of a Schwarzshild black
hole. When the new equation of state density is utilized to
investigate the entropy of a scalar field outside the horizon of a
static black hole, the divergence appearing in the brick wall
model is removed, without any cutoff. The entropy proportional to
the horizon area is derived from the contribution of the vicinity
of the horizon.

\end{abstract}
\newpage
 The title is the same as Ref.\cite{dlm} where Demers et al show
that the divergence appearing in the brick wall model\cite{thooft}
can be absorbed into the renormalized Newton's constant. By using
the WKB approximation, 't Hooft investigates the statistical
properties of a scalar field outside the horizon of a
Schwarzschild black hole. The entropy proportional to the horizon
area is obtained, but with a cutoff utilized to remove the
divergence of the density of states. The cutoff is introduced by
hand and looks  unnatural. Susskind and Uglum suggest that the
explosive free energy and  entropy in the model of 't Hooft are
related to the divergence of the one-loop effective action of the
quantum field theory in curved space\cite{ug}. Their conjecture is
confirmed by \cite{dlm}. The authors of \cite{dlm} remove the
cutoff and regularize the divergent free energy and entropy by
introducing some regulators. These fictitious fields are
especially designated in the number, statistics and masses. To my
surprise, the entropy expressed by the masses of the regulators
can be precisely renormalized to the Bekenstein-Hawking formula,
$S=A/(4G_{R})$, $G_{R}$ is the renormalized Newton's constant.
However, it is hard to understand the introduction of the `` bare
entropy" in Ref.\cite{dlm}. The `` bare entropy" seems to be
negative and its meaning is unclear\footnote{Dr. Fursaev told me,
this difficulty can be overcome in the Sakharov's induced
gravity\cite{fusaev}.}. Is there a better method can remove
 the divergence appearing in the brick wall model?

Recently, many efforts have been devoted to the generalized
uncertainty relation
\begin{eqnarray}\label{uncert}
\Delta x\Delta p\geq \hbar+\frac{\lambda}{\hbar}(\Delta p)^2,
\end{eqnarray}
and its consequences\cite{kempf}--\cite{chang}, especially the
effect on the density of states\cite{rama}\cite{chang}. Here
$\hbar$ is the Planck constant, $\lambda$ is of order of the
Planck length. Eq. (\ref{uncert}) means that there is a minimal
length, $2\sqrt{\lambda}$. As well known, the number of quantum
states in the integrals $d^3{\bf x}d^3{\bf p}$ is given by
\begin{eqnarray}\label{numb}
\frac{d^3{\bf x}d^3{\bf p}}{(2\pi \hbar)^3},
\end{eqnarray}
which can be understood as follows: since the uncertainty relation
$\Delta x \Delta p \sim 2\pi \hbar$,  one quantum state
corresponds to a ``cell" of volume $(2\pi \hbar)^3$ in the phase
space. Based on the Liuville theorem, the authors of
Ref.\cite{chang} argue that the number of quantum states should be
changed to the following
\begin{eqnarray}\label{num}
\frac{d^3{\bf x}d^3{\bf p}}{(2\pi \hbar)^3(1+\lambda p^2)^3},
\end{eqnarray}
where $p^2=p_{i} p^{i},i=1, 2, 3$. Eq. (\ref{num}) seriously
deforms the Planckian spectrum of the black body radiation at the
Planck temperature, $T_{\lambda}=\sqrt{1/\lambda}$ (see
Ref.\cite{chang}, fig.2).

Let us discuss the more details than Ref.\cite{chang}. This will
benefit the following investigation of the black hole entropy.
From Eq.(\ref{num}), we directly write down the density of
internal energy of the thermal radiation
\begin{eqnarray}
u&=&\int_{0}^{\infty} \frac{\omega^3 d\omega}{(e^{\beta \omega}-1)(1+\lambda
\omega^2)^3}\nonumber\\
&=&\beta^{-4}\int_{0}^{\infty} \frac{x^3
dx}{(e^{x}-1)(1+ax^2)^3}\nonumber\\
&=&\beta^{-4}G(a),
\end{eqnarray}
where $a=\lambda/\beta^2, x=\beta\omega$. We take the units
$G=c=\hbar=k_B=1$.
 The above integral can not be expressed
as a simple formula, but we can investigate its asymptotic
behavior in the two different conditions. We first consider the
case $a\ll 1$. This means that the temperature is much
less than the Planck temperature. We have
\begin{eqnarray}
G(0)&=&\int_{0}^{\infty} \frac{x^3 dx}{e^{x}-1}=\frac{\pi^4}{15},\nonumber\\
G^{\prime}(0)&=&-3\int_{0}^{\infty} \frac{x^5 dx}{e^{x}-1}=-\frac{24\pi^6}{63},
\end{eqnarray}
then
\begin{eqnarray}\label{nnmd}
u&=&\beta^{-4}[G(0)+G^{\prime}(0)a]\nonumber\\
&=&\frac{\pi^4}{15}\beta^{-4}\left(1-\frac{40\pi^2\lambda}{7\beta^2}\right).
\end{eqnarray}
In the usual case, above equation does not essentially change the
well known conclusion for the black body radiation because the
correction is very slight. For example, the temperature of the
center of the neutron star is $10^{9}K$, but the Planck
temperature is $10^{32}K, \lambda/\beta^2\sim 10^{-46}$. However,
Eq. (\ref{nnmd}) is no longer valid for the case
$\lambda/\beta^2\gg 1$, that is higher than the Planck
temperature. We calculate the upper bound of energy density, that
is
\begin{eqnarray}\label{bound}
u&<&\beta^{-4}\int_{0}^{\infty} \frac{x^2 dx}{(1+\frac{\lambda
x^2}{\beta^2})^3}\nonumber\\
&=&\beta^{-4}\cdot\frac{\pi}{16}\left(\frac{\lambda}
{\beta^2}\right)^{-3/2}\nonumber\\
&=&\frac{\pi}{16\lambda^{3/2}}\beta^{-1},
\end{eqnarray}
 where the inequality is due to $e^x-1>x$. This means that when the temperature is higher than the Planck
temperature the state equation of the thermal radiation is
essentially different from the well known conclusion, $u\sim
\beta^{-4}$. This will influence the emission of the black hole.
According to the Stefan-Boltzmann law, the loss mass rate of a
Schwarzscild black hole reads
\begin{eqnarray}\label{dm}
\frac{dM}{dt}\sim \beta^{-4}A\sim \frac{1}{M^2},
\end{eqnarray}
where $M$ the mass of the hole. At the last stage of emission,
$M\rightarrow 0$, so the emission rate becomes divergent. However,
from Eq. (\ref{bound}), at the last stage, the rate will be
changed to
\begin{eqnarray}
\frac{dM}{dt}\sim \beta^{-1} A\sim M\rightarrow 0,
\end{eqnarray}
here is no burst.

We turn to the problem of black hole entropy. Recalling the brick
wall model, the number of quantum states less than energy $\omega$
is given by\cite{thooft}\cite{padhan}\cite{jing}
\begin{eqnarray}\label{gama}
\Gamma(\omega)=\frac{2\omega^3}{3\pi}\int_{r_0+\epsilon}^{L}\frac{dr}{f^2},
\end{eqnarray}
which is for a massless scalar field in a spherical and static
space-time as follows
\begin{eqnarray}
ds^2=-fdt^2+f^{-1}dr^2+r^2d\theta^2+r^2\sin^2\theta d\phi^2,
\end{eqnarray}
where $f=f(r)$. The horizon is located by $f(r_0)=0$.
 $\epsilon$ is the cutoff near the horizon. Obviously, the number
 of states is divergent if $\epsilon=0$. We carefully check the
 derivation of Eq. (\ref{gama}) and find that it agrees with Eq.
 (\ref{numb}), not (\ref{num}). The former leads to the following
 formula
\begin{eqnarray}\label{usual}
S=\frac{8\pi^3}{45\beta^3}\int\frac{r^2dr}{f^2},
\end{eqnarray}
which is analogous with the usual state equation of the thermal
radiation: $(\beta\sqrt{f})^{-1}$ is the local temperature, $4\pi
r^2dr/\sqrt{f}$ is the element of the spatial volume of the
spherical shell. The divergent entropy means the invalidity of the
usual state equation near the black hole horizon. If we take Eq.
(\ref{num}), the
 situation may be essentially different. Why not have a try?
 Substituting the wave function $\Phi=\exp(-i\omega t)\psi(r, \theta,
 \varphi)$ into the equation of massless scalar field
 \begin{eqnarray}
\frac 1{\sqrt{-g}}\partial _\mu (\sqrt{-g}g^{\mu \nu }\partial
_\nu \Phi )=0,
\end{eqnarray}
we obtain
\begin{eqnarray}
\frac{\partial^2\psi}{\partial
r^2}+\Big(\frac{f^{\prime}}{f}+\frac{2}{r}\Big)
\frac{\partial\psi}{\partial r}+\frac{1}{f}
\left[\frac{\omega^2}{f}+\frac{1}{r^2}\left(\frac{\partial^2}{\partial\theta^2}+
\cot\theta\frac{\partial}{
\partial\theta}+\frac{1}{\sin^2 \theta}\frac{\partial^2}{\partial\varphi^2}%
\right)\right]\psi=0.\nonumber\\
\end{eqnarray}
By using the WKB approximation with $\psi\sim \exp[iS(r, \theta,
\phi)]$, we have
\begin{eqnarray}
p_r^2=\frac{1}{f}\left[\frac{\omega^2}{f}-\frac{1}{r^2}p_{\theta}^2-\frac{1}
{r^2\sin^2\theta}p_{\varphi}^2\right],
\end{eqnarray}
where
\begin{eqnarray}
p_{r}=\frac{\partial S}{\partial r}, p_{\theta}=\frac{\partial
S}{\partial \theta}, p_{\varphi}=\frac{\partial S}{\partial
\varphi}.
\end{eqnarray}
 We also obtain the square module of momentum
\begin{eqnarray}
p^2=p_{i}p^{i}=g^{rr}p_{r}^2+g^{\theta\theta}p_{\theta}^2+
g^{\varphi\varphi}p_{\varphi}^2=\frac{\omega^2}{f}.
\end{eqnarray}
From Eq. (\ref{num}), the number of quantum states with energy
less than $\omega$ is given by
\begin{eqnarray}\label{gom}
g(\omega)&=&\frac{1}{(2\pi)^3}\int \frac{drd\theta d\varphi dp_r
dp_{\theta}dp_{\varphi}}{(1+\lambda \omega^2/f)^3}\nonumber\\
&=&\frac{1}{(2\pi)^3}\int \frac{drd\theta d\varphi}{(1+\lambda\omega
^2/f)^3}\int\frac{2}{f^{1/2}}\left[\frac{\omega^2}{f}-\frac{1}{r^2}p_{\theta}^2-\frac{1}
{r^2\sin^2\theta}p_{\varphi}^2\right]^{1/2}dp_{\theta}dp_{\varphi}\nonumber\\
&=&\frac{4\pi\omega^3}{3(2\pi)^3}\int \frac{r^2dr}{f^2(1+\lambda
\omega^2/f)^3}\int \sin\theta d\theta d\varphi\nonumber\\
&=&\frac{2\omega^3}{3\pi}\int \frac{r^2dr}{f^2(1+\lambda
\omega^2/f)^3},
\end{eqnarray}
where the integration goes over those values of $p_{\theta},
p_{\varphi}$ for which the argument of the square root is
positive(please refer to Refs.\cite{thooft},\cite{padhan} and
\cite{jing}). When $\lambda=0$, Eq. (\ref{gom}) naturally returns
to (\ref{gama}). However, in the case $\lambda\neq 0$, Eq.
(\ref{gom}) is essentially different from (\ref{gama}): it is
convergent at the horizon without any cutoff! By using the usual
method, the free energy is given by
\begin{eqnarray}
F(\beta)&=&\frac{1}{\beta}\int
dg(\omega)\ln(1-e^{-\beta\omega})\nonumber\\
&=&-\int_{0}^{\infty} \frac{g(\omega)d\omega}{e^{\beta
\omega}-1}\nonumber\\
&=&-\frac{2}{3\pi}\int_{r_0}\frac{r^2dr}{f^2}\int_{0}^{\infty}\frac{\omega^3d\omega}
{(e^{\beta \omega}-1)(1+\lambda\omega^2/f)^3}.
\end{eqnarray}
The entropy reads
\begin{eqnarray}
S&=&\beta^2\frac{\partial F}{\partial \beta}\nonumber\\
&=&\frac{2\beta^2}{3\pi}\int_{r_0}\frac{r^2dr}{f^2}\int_{0}^{\infty}\frac{e^{\beta\omega}
\omega^4d\omega}
{(e^{\beta \omega}-1)^2(1+\lambda\omega^2/f)^3}\nonumber\\
&=&\frac{2\beta^{-3}}{3\pi}\int_{r_0}\frac{r^2dr}{f^2}\int_{0}^{\infty}\frac{x^4dx}
{(1-e^{-x})(e^{x}-1)(1+\frac{\lambda x^2}{\beta^2 f})^3},
\end{eqnarray}
where $x=\beta\omega$. Taking into account the following
inequalities
\begin{eqnarray}
1-e^{-x}&>&\frac{x}{1+x}, \nonumber\\
e^{x}-1&>&x,
\end{eqnarray}
We obtain
\begin{eqnarray}
S&<&\frac{2\beta^{-3}}{3\pi}\int_{r_0}\frac{r^2dr}{f^2}\int_{0}^{\infty}\frac{(x^3+x^2)dx}
{(1+\frac{\lambda x^2}{\beta^2 f})^3}\nonumber\\
&=&\frac{2\beta^{-3}}{3\pi}\int_{r_0}\frac{r^2dr}{f^2}\left[\frac{1}{4}
\left(\frac{\lambda}{\beta^2f}\right)^{-2}+\frac{\pi}{16}
\left(\frac{\lambda}{\beta^2f}\right)^{-3/2}\right]\nonumber\\
&=&\frac{\beta}{6\pi\lambda^2}\int_{r_0}r^2dr+\frac{\lambda^{-3/2}}{24}
\int_{r_0}\frac{r^2dr}{f^{1/2}}.
\end{eqnarray}
We are only interested in the contribution from the vicinity near
the horizon, $[r_0, r_0+\epsilon]$, which corresponds to a proper
distance of order of the minimal length, $2\sqrt{\lambda}$. This
is because the entropy closes to the upper bound only in this
vicinity. Furthermore, it is just the vicinity neglected by brick
wall model. We have
\begin{eqnarray}
2\sqrt{\lambda}&=&\int_{r_0}^{r_0+\epsilon}\frac{dr}{\sqrt{f}}\nonumber\\
&\approx & \int_{r_0}^{r_0+\epsilon}\frac{dr}{\sqrt{2\kappa
(r-r_0)}}\nonumber\\
&=&\sqrt{\frac{2\epsilon}{\kappa}},
\end{eqnarray}
where $\kappa$ is the surface gravity at the horizon of black hole
and it is identified as $\kappa=2\pi\beta^{-1}$.
 Thus we naturally derive the entropy proportional to the
horizon area
\begin{eqnarray}
S&\sim&
\frac{\beta}{6\pi\lambda^2}r_{0}^2\epsilon+\frac{\lambda^{-3/2}}{24}\cdot
2r_0^2\sqrt{\lambda}\nonumber\\
&=&\frac{3A}{16\lambda\pi},
\end{eqnarray}
where $A=4\pi r_{0}^2$ is the surface area of the black hole.

As early as 1992, Li and Liu phenomenally proposed that the state
equations of the thermal radiation near the horizon should be
changed to a series of new formulae rather than Eq. (\ref{usual}),
in order to maintain the validity of the generalized second law of
thermodynamics\cite{liliu}. Using the Li-Liu equation, Wang
investigates the entropy of a self-gravitational radiation system
and obtains the Bekenstein's entropy bound\cite{wang}. Here,
Parallel to the brick wall model, the scalar field near the
horizon of a static black hole is investigated again, we obtain
the entropy proportional to the horizon area. There is no
divergence
 without any cutoff near the horizon. This convergency is due to the effect
 of the generalized uncertainty relation on the quantum states. This provides
an evidence for the idea of Li and Liu. The more details between the Li-Liu
equation and the generalized uncertainty relation will be investigated
in the future.

As pointed by Ref.\cite{garay}, the generalized uncertainty
relation (\ref{uncert}) may have a dynamical origin since it
contains a dimensional coupling constant $\lambda$. If Eq.
(\ref{uncert}) is indeed due to the string theory, $\lambda$
should be associated with the stringy scale $l_s^2$. This implies
that one takes into account the contribution from the stringy
excitation when calculating the density of quantum states. The
convergent property should be reexamined. There are some
 evidences (or arguments) for the convergence of the density of
 states even if considering the stringy excitation: firstly, the
 bosonic string can be described by a discrete field theory, then
 the number of degrees of freedom of it is smaller than that of
 the usual field theory\cite{kleb}; Secondly, The entropy of a string is
 proportional to its mass since the degeneracy increases exponentially with
 the mass level. However, the massive string can not be excited in
 the low energy effective theory (such as general relativity)\cite{peet}.
Therefore, the contribution from the stringy excitation is
ignored in the case of the massive black hole where the semi-classical
approximation is still valid. As to the black hole at the Planck
scale the usual quantum field theory is no longer valid. the
entropies of the black hole and the excited string states are
matched in the correspondence principle\cite{corres}.
\section*{Acknowledgements}
The author thanks X. J.Wang, Z. Q. Bai and H. G. Luo for their
zealous helps during the research. The author also thank Profs. Y.
Z. Zhang, R. G. Cai and C. G. Huang for their comments on this
research. This work is supported by the Post Doctor Foundation of
China and K. C. Wong Education Foundation, Hong Kong.

\end{document}